\documentclass[dvips]{PoS}
\usepackage{amsmath,amssymb}
\usepackage{graphicx}
\usepackage{subfigure}

\title{$\eta$ and $\eta'$ meson masses from $N_f=2+1+1$ twisted mass
  lattice QCD}

\ShortTitle{$\eta$ and $\eta'$ masses from $N_f=2+1+1$ tmLQCD}

\author{\speaker{Konstantin Ottnad}, Carsten Urbach\\
  Helmholtz Institut f{\"u}r Strahlen und Kernphysik and
  Bethe Center for Theoretical Physics,\\
  Universit{\"a}t Bonn, Nussallee 14-16, 53115 Bonn, Germany\\
  E-mail: \email{ottnad,urbach@hiskp.uni-bonn.de}
}

\author{Chris Michael\\
  Theoretical Physics Division, Department of Mathematical Sciences\\
  The University of Liverpool, Liverpool, L69 3BX, UK\\
  E-mail: \email{c.michael@liv.ac.uk}
}

\author{Siebren Reker\\
  Centre for Theoretical Physics\\
  University of Groningen, Nijenborgh 4, 9747 AG Groningen, The Netherlands\\
  E-mail: \email{s.f.reker@rug.nl}
}

\author{for the European Twisted Mass collaboration}

\abstract{
  We determine mass and flavour content of $\eta$ and $\eta'$
  states using $N_f=2+1+1$ Wilson twisted mass lattice QCD. We
  describe how those flavour singlet states need to be treated in this
  lattice formulation. Results are presented for two values of 
  the lattice spacing, $a\approx0.08\ \mathrm{fm}$ and $a\approx0.09\
  \mathrm{fm}$, with a range of light quark masses corresponding to
  values of the pion mass from $270$ to $500\ \mathrm{MeV}$
  and fixed bare strange and charm quark mass values.
}

\FullConference{The XXIX International Symposium on Lattice Field Theory, Lattice2011\\
		July 11-16, 2011\\
		Lake Tahoe, USA}

\begin{document}

\section{Introduction}

From experiment it is known that the masses of the nine light
pseudo-scalar
mesons show an interesting pattern. Taking the quark model point of
view, the three lightest mesons, the pions, contain only the two
lightest quark flavours, the \emph{up}- and \emph{down}-quarks. The
pion triplet has a mass of $M_\pi\approx140\ \mathrm{MeV}$. For the other six,
the \emph{strange} quark contributes also, and hence they are  
heavier. In contrast to what one might expect  five of them, 
the four kaons and the $\eta$ meson,  have        
roughly equal mass around $500$ to $600\ \mathrm{MeV}$, while
the last one, the $\eta'$ meson, is much
heavier, with mass of about $1\ \mathrm{GeV}$. On the QCD level, the
reason for this pattern is thought to be the breaking of the $U_A(1)$
symmetry by quantum effects. The $\eta'$ meson is, even in a
world with three massless quarks, not a Goldstone boson. 
In this proceeding contribution, we discuss the determination of $\eta$
and $\eta'$ meson masses using twisted mass lattice QCD (tmLQCD) with
$N_f=2+1+1$ dynamical quark flavours. This will not only allow a 
study of the dependence of the $\eta,\eta'$ masses on the light quark mass
value, but also an investigation of the charm quark contribution to
both of these states. Moreover, the $\eta_c$ meson mass can be studied
in principle. For recent lattice studies in $N_f=2+1$ flavour QCD
see~\cite{Christ:2010dd,Kaneko:2009za}. 

\begin{table}[t!]
  \centering
  \begin{tabular*}{.8\textwidth}{@{\extracolsep{\fill}}lccccc}
    \hline\hline
    ensemble & $\beta$ & $a\mu_\ell$ & $a\mu_\sigma$ & $a\mu_\delta$ &
    $L/a$ \\
    \hline\hline
    A40.24  & $1.90$ & $0.0040$ & $0.150$ & $0.190$ & $24$ \\
    A60.24  & $1.90$ & $0.0060$ & $0.150$ & $0.190$ & $24$ \\
    A80.24  & $1.90$ & $0.0080$ & $0.150$ & $0.190$ & $24$ \\
    \hline
    A80.24s & $1.90$ & $0.0080$ & $0.150$ & $0.197$ & $24$ \\
    \hline
    B25.32 & $1.95$ & $0.0025$ & $0.135$ & $0.170$ & $32$ \\ 
    B35.32 & $1.95$ & $0.0035$ & $0.135$ & $0.170$ & $32$ \\ 
    B85.24 & $1.95$ & $0.0085$ & $0.135$ & $0.170$ & $24$ \\
    \hline\hline
  \end{tabular*}
  \caption{The ensembles used in this investigation. The notation of
    ref.~\cite{Baron:2010bv} is used for labeling the ensembles.}
  \label{tab:setup}
\end{table}

\section{Lattice Action}

We use gauge configurations as produced by the European Twisted Mass
Collaboration (ETMC) with $N_f=2+1+1$ flavours of Wilson twisted mass
quarks and Iwasaki gauge action~\cite{Baron:2010bv,Baron:2011sf}. The
details are described in ref.~\cite{Baron:2010bv} and the ensembles 
used in this investigation 
are summarised in table~\ref{tab:setup}. The twisted mass Dirac
operator in the light -- i.e. up/down -- sector
reads~\cite{Frezzotti:2000nk}
\begin{equation}
  \label{eq:Dud}
  D_\ell = D_W + m_0 + i \mu_\ell \gamma_5\tau^3
\end{equation}
and in the strange/charm sector~\cite{Frezzotti:2003xj}
\begin{equation}
  \label{eq:Dsc}
  D_h = D_W + m_0 + i \mu_\sigma \gamma_5\tau^1 + \mu_\delta \tau^3\, ,
\end{equation}
where $D_W$ is the Wilson Dirac operator. The value of $m_0$ was tuned
to its critical value as discussed in
refs.~\cite{Chiarappa:2006ae,Baron:2010bv} in 
order to realise automatic $\mathcal{O}(a)$ improvement at maximal
twist~\cite{Frezzotti:2003ni}. Note that the bare twisted masses
$\mu_{\sigma,\delta}$ are related to the bare strange and charm quark
masses via the relation
\begin{equation}
  \label{eq:msc}
  m_{c,s} = \mu_\sigma\ \pm\ (Z_\mathrm{P}/Z_\mathrm{S})\ \mu_\delta
\end{equation}
with pseudo-scalar and scalar renormalisation constants $Z_\mathrm{P}$
and $Z_\mathrm{S}$. Quark fields in the twisted basis are denoted by 
$\chi_{\ell,h}$ and in the physical basis by
$\psi_{\ell,h}$. They are related via the axial rotations
\begin{equation}
  \chi_\ell = e^{i\pi\gamma_5\tau^3/4}\psi_\ell\,,\quad\bar\chi_\ell
  = \bar\psi_\ell\ e^{i\pi\gamma_5\tau^3/4}\,,\qquad
  \chi_h = e^{i\pi\gamma_5\tau^1/4}\psi_h\,,\quad\bar\chi_h
  = \bar\psi_h\ e^{i\pi\gamma_5\tau^1/4}\,.
\end{equation}
With automatic $\mathcal{O}(a)$ improvement being the biggest
advantage of tmLQCD at maximal twist, the downside is that flavour
symmetry is broken at finite values of the lattice spacing. This was
shown to affect mainly the mass value of the neutral pion
mass~\cite{Urbach:2007rt,Dimopoulos:2009qv,Baron:2009wt}, however, in
the case of $N_f=2+1+1$ dynamical quarks, it implies the complication of
mixing between strange and charm quarks.  

\section{Flavour Singlet Pseudo-Scalar Mesons in $N_f=2+1+1$ tmLQCD}

In order to compute masses of pseudo-scalar flavour singlet mesons we
have to include light, strange and charm contributions to build the
appropriate correlation functions. In the light sector, one appropriate
operator is given by~\cite{Jansen:2008wv}
\begin{equation}
  \label{eq:ll}
  \frac{1}{\sqrt{2}}(\psi_u i \gamma_5\psi_u + \psi_d i \gamma_5\psi_d)
  \quad \to\quad \frac{1}{\sqrt{2}}(\bar\chi_u \chi_u - \bar\chi_d \chi_d)\equiv\ell\, .
\end{equation}
In the strange and charm sector, the corresponding operator reads
\begin{equation}
  \label{eq:physsinglet}
  \begin{pmatrix}
    \bar \psi_c \\
    \bar \psi_s \\
  \end{pmatrix}
  i \gamma_5 \frac{1 \pm \tau^3}{2}
  \begin{pmatrix}
    \psi_c \\
    \psi_s \\
  \end{pmatrix}
  \quad\to\quad
  \begin{pmatrix}
    \bar \chi_c \\
    \bar \chi_s \\
  \end{pmatrix}
  \frac{-\tau^1  \pm  i \gamma_5\tau^3}{2}
  \begin{pmatrix}
    \chi_c \\
    \chi_s \\
  \end{pmatrix}\, .
\end{equation}
In practice we need to compute correlation functions of the
following interpolating operators
\begin{equation}
  \label{eq:psequ}
  \begin{split}
    P_{ss}\ \equiv\ (\bar\psi_s i \gamma_5\psi_s) &=
    (\bar\chi_c i \gamma_5\chi_c - \bar\chi_s i \gamma_5\chi_s)/2 -
    \frac{Z_\mathrm{S}}{Z_\mathrm{P}}(\bar\chi_s\chi_c +
    \bar\chi_c\chi_s)/2\, ,  \\
    P_{cc}\ \equiv\ (\bar\psi_c i \gamma_5\psi_c) &=
    (\bar\chi_s i \gamma_5\chi_s - \bar\chi_c i \gamma_5\chi_c)/2 -
    \frac{Z_\mathrm{S}}{Z_\mathrm{P}}(\bar\chi_s\chi_c +
    \bar\chi_c\chi_s)/2\, .  \\
  \end{split}
\end{equation}
Note that the sum of pseudo-scalar and scalar contributions
appears with the ratio of renormalisation factors $Z\equiv
Z_\mathrm{S}/Z_\mathrm{P}$, which needs to be  
taken into account properly. $Z$ has not yet been determined for
all values of $\beta$ non-perturbatively. 

However, for the mass determination, we can avoid this complication 
by changing the basis and compute the real and positive 
definite correlation matrix
\begin{equation}
  \label{eq:corrmatrix}
  \mathcal{C} = 
  \begin{pmatrix}
    \eta_{\ell\ell} & \eta_{\ell P_h} & \eta_{\ell S_h} \\
    \eta_{P_h \ell} & \eta_{P_h P_h} & \eta_{P_h S_h} \\
    \eta_{S_h \ell} & \eta_{S_h P_h} & \eta_{S_h S_h} \\
  \end{pmatrix}\, ,
\end{equation}
with the notation 
\begin{equation}
  P_h \equiv (\bar\chi_c i \gamma_5\chi_c - \bar\chi_s i
  \gamma_5\chi_s)/2\, ,\qquad
  S_h \equiv (\bar\chi_s\chi_c + \bar\chi_c\chi_s)/2
\end{equation}
and $\eta_{XY}$ denoting the corresponding correlation
function. Masses can determined by solving the generalised eigenvalue 
problem~\cite{Michael:1982gb,Luscher:1990ck}
\begin{equation}
  \mathcal{C}(t)\ \eta^{(n)}(t,t_0) = \lambda^{(n)}(t, t_0)\ \mathcal{C}(t_0)\
  \eta^{(n)}(t,t_0) \, . 
\end{equation}
Taking into account the periodic boundary conditions for a meson, we
can determine the effective masses by solving
\begin{equation}
  \frac{\lambda^{(n)}(t,t_0)}{\lambda^{(n)}(t+1,t_0)} = \frac{e^{-m^{(n)}
      t}+ e^{-m^{(n)}(T-t)}}
  {e^{-m^{(n)}(t+1)}+ e^{-m^{(n)}(T-(t+1))}}
\end{equation}
for $m^{(n)}$, where $n$ counts the eigenvalues. The state with the
lowest mass should correspond to the $\eta$ and the second state
to the $\eta'$ meson.

From the components $\eta^{(n)}_{0,1,2}$ of the eigenvectors, we can
reconstruct the physical flavour contents $c^{(n)}_{\ell,s,c}$ from
\begin{equation}
  \begin{split}
    c_\ell^{(n)} &= \frac{1}{\mathcal{N}^{(n)}} (\eta^{(n)}_0)\\
    c_s^{(n)} &= \frac{1}{\mathcal{N}^{(n)}} (- Z \eta^{(n)}_1 +
    \eta^{(n)}_2)/\sqrt{2}\\ 
    c_c^{(n)} &= \frac{1}{\mathcal{N}^{(n)}} (- Z \eta^{(n)}_1 -
    \eta^{(n)}_2)/\sqrt{2}\\ 
  \end{split}
\end{equation}
with normalisation
\[
\mathcal{N}^{(n)} = \sqrt{(\eta^{(n)}_0)^2 +  (Z \eta^{(n)}_1)^2 + (\eta^{(n)}_2)^2}\ .
\]
At this point the ratio $Z\equiv Z_\mathrm{S}/Z_\mathrm{P}$ is needed
again. Assuming for a moment that  charm does not contribute  
significantly to the $\eta$ and $\eta'$ states, one can extract the
$\eta$-$\eta'$ mixing angle $\phi$ from
\begin{equation}
  \label{eq:angle}
  \cos(\phi) = c_\ell^{(0)} \approx c_s^{(1)},\quad sin(\phi) =
  -c_s^{(0)} = c_\ell^{(1)}
\end{equation}
with $^{(0)}$ ($^{(1)}$) denoting the $\eta$ ($\eta'$) state.

\section{Results}

We have computed all contractions needed for building the correlation 
matrix of eq.~(\ref{eq:corrmatrix}). For the connected contributions, we 
used stochastic time-slice sources (the so called
``one-end-trick''~\cite{Boucaud:2008xu}). For the disconnected
contributions, we used stochastic volume sources with complex
Gaussian noise~\cite{Boucaud:2008xu}. As discussed in
ref.~\cite{Jansen:2008wv} one can estimate the light disconnected
contributions very efficiently using the identity
\[
D_u^{-1} - D_d^{-1} = -2i\mu_\ell D_d^{-1}\ \gamma_5\ D_u^{-1}\ .
\]
For the heavy sector such a simple relation does not exist, but we
can use the so called hopping parameter variance reduction, which
relies on the same equality as in the mass degenerate two flavour
case (see ref.~\cite{Boucaud:2008xu} and references therein)
\[
D_h^{-1} = B - BHB + B(HB)^2 - B(HB)^3 + D_h^{-1} (HB)^4
\]
with $D_h = (1 + HB) A$, $B = 1/A$ and $H$ the two flavour hopping
matrix. We use $24$ stochastic volume sources per gauge
configuration in both the heavy and the light sector. 

We use both local and fuzzed sources to enlarge our correlation  
matrix by a factor two. In addition to the interpolating operator
quoted in eqs.~(\ref{eq:ll}) and (\ref{eq:physsinglet}), we also
plan to consider the $\gamma$-matrix combination $i\gamma_0\gamma_5$, which
will increase the correlation matrix by another factor of two. The number of 
gauge configurations investigated per ensemble is in most cases 
around $1200$, and for ensemble B25.32 is $1500$. Statistical errors 
are computed using the bootstrap method with $1000$ samples.

\begin{figure}[t]
  \centering
  \subfigure[]{\includegraphics[width=.48\linewidth]
  {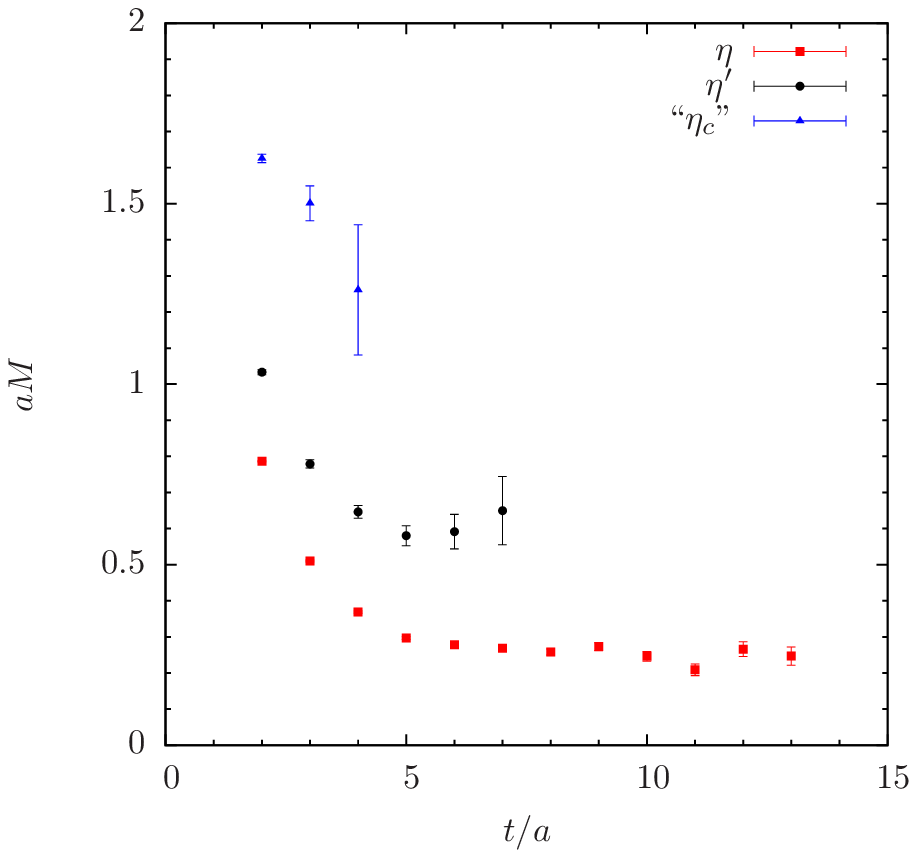}}\quad
  \subfigure[]{\includegraphics[width=.48\linewidth]
  {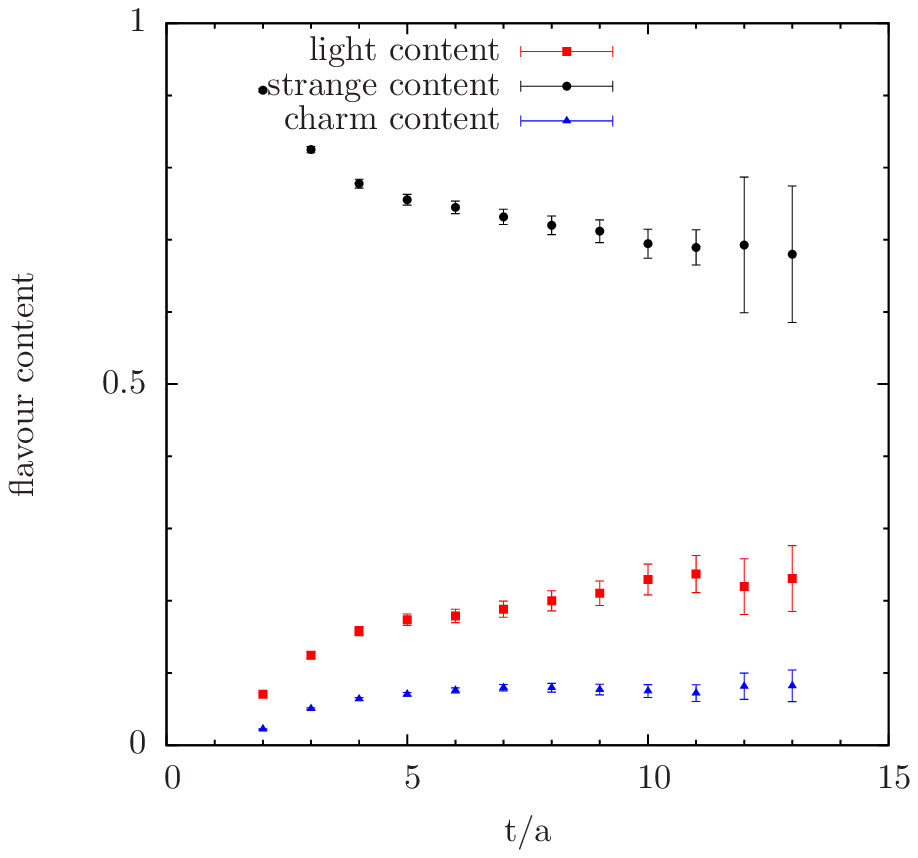}}
  \caption{(a) Effective masses in lattice units determined from solving
    the generalised eigenvalue problem with $t_0/a=1$ for ensemble
    B25.32. We show the results extracted from a $3\times3$
    matrix. (b) squared flavour content of $\eta$ for B25.32.}   
  \label{fig:effmasses}
\end{figure}

In figure~\ref{fig:effmasses} we show the effective masses determined
from solving the generalised eigenvalue problem for ensemble
B25.32 from a $3\times3$ matrix with local operators only. We kept
$t_0/a=1$ fixed. One observes that the ground state is very well
determined and it 
can be extracted from a plateau fit. The second state, i.e. the    
$\eta'$, is much more noisy and a mass determination is questionable,
at least from a $3\times3$ matrix. Enlarging the matrix size 
significantly reduces the contributions of excited states to the
lowest states and, due to smaller statistical errors at smaller $t$
values, a determination becomes possible. The third state appears to be
in the region where one would expect the $\eta_c$ mass value, however,
the signal is lost at $t/a=5$ already, which makes a reliable
determination not feasible.  

\begin{figure}[t]
  \centering
  \includegraphics[width=.7\linewidth]%
  {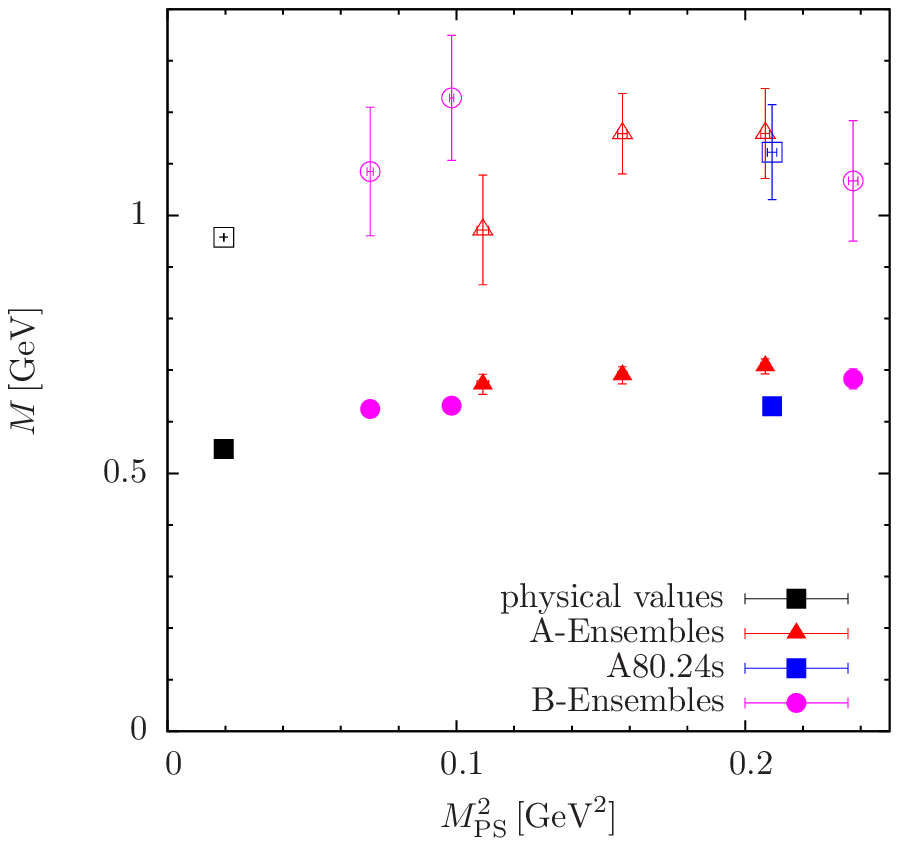}
  \caption{Preliminary values for $m_\eta$ and $m_{\eta'}$ for two
    $\beta$-values in physical units 
    as a function of the squared pion mass. Filled symbols represent
    $m_\eta$, open ones $m_{\eta'}$.}
  \label{fig:masssummary}
\end{figure}

In figure~\ref{fig:masssummary} we show the masses of the $\eta$ and
$\eta'$ mesons for the various ensembles we used as a function of the
squared pion mass. In addition we show the corresponding physical
values. The scale was set from $f_\pi$ and $m_\pi$ using the results of
ref.~\cite{Baron:2010bv}. It is clear that the $\eta$ meson 
mass can be extracted with high precision, while the $\eta'$ meson
mass requires a larger correlation matrix, which is work in
progress. The comparison with the corresponding physical values
seems to point towards good agreement. 

We also determine the flavour content of the two states as explained
above. It turns out that the $\eta$ has a dominant strange quark
content (see right panel of figure~\ref{fig:effmasses}), while the
$\eta'$ is dominated by light quarks. For both the  
charm contribution is rather small, however, for the $\eta$ it turns
out to be significantly non-zero. A preliminary determination of the 
mixing angle eq.~(\ref{eq:angle}) yields a very stable value of about 
$60\,^{\circ}$. 
Note that this is the mixing angle to the flavour
eigenstates. The computation of the mixing angle with respect to the
$\eta_0$ and $\eta_8$ states is in progress.

\subsection{Bare Strange and Charm Quark Mass Dependence}

The results displayed in figure~\ref{fig:masssummary} have been
obtained using the bare values of $\mu_\sigma$ and $\mu_\delta$ as
used for the production of the ensembles. Those values, however, did
not lead to the correct values of, for e.g., the kaon and $D$-meson masses, 
see e.g. ref.~\cite{Baron:2011sf}. Moreover, the physical strange
and charm quark mass values differ between the $A$ and $B$       
ensembles. Hence, figure~\ref{fig:masssummary} is not yet conclusive
with regards to the size of lattice artifacts and extrapolation to the 
physical point. What we can learn is that the light quark mass
dependence in both states appears to be rather weak. 

We also have two ensembles A80.24 and A80.24s with different bare
values for $\mu_{\sigma,\delta}$ and a retuned value for $\kappa$ 
but identical parameters  
otherwise. Ensemble A80.24s is significantly better 
tuned with respect to the physical kaon mass
value~\cite{Baron:2011sf}. As seen in figure~\ref{fig:masssummary}, 
our results seem to indicate that this change in the bare parameters
also has a significant impact on the $\eta$ meson mass value, while the
$\eta'$ mass is unaffected within the (rather large) errors. 

\section{Summary and Outlook}

We presented a computation of $\eta$ and $\eta'$ meson masses from
$N_f=2+1+1$ Wilson twisted mass lattice QCD. The results we obtained
so far are rather encouraging, the $\eta$ meson mass can be determined
with high precision. Also for the $\eta'$ meson mass, we hope to be
able to give more precise results by increasing the correlation matrix
under investigation. 

As it is not so easy to tune bare strange and charm quark masses
exactly, we shall in the future use a mixed action approach for strange
and charm quarks. This will not only avoid the complication induced by
flavour symmetry breaking in the twisted mass formulation, but it will
also allow to use the efficient noise reduction techniques in the
heavy sector. However, this method requires a careful investigation of
unitarity breaking effects.

\subsection*{Acknowledgments}  

We thank J.~Daldrop, K.~Jansen, C.~McNeile, M.~Petschlies,
M.~Wagner and F.~Zimmermann for useful discussions. We thank the
members of ETMC  for the most enjoyable collaboration. The computer
time for this project was made available to us by the John von
Neumann-Institute for Computing (NIC) on the JUDGE and Jugene systems
in J{\"u}lich. This project was funded by the DFG as a
project in the SFB/TR 16. The open source software packages
tmLQCD~\cite{Jansen:2009xp} and Lemon~\cite{Deuzeman:2011wz} have been
used.

\bibliographystyle{h-physrev5}
\bibliography{bibliography}

\end{document}